\documentclass[twocolumn,aps,floatfix]{revtex4}
\usepackage{amsmath}
\usepackage{amssymb}
\usepackage{graphicx}
\usepackage{color}

\newcommand{\be}{\begin{equation}}
\newcommand{\ee}{\end{equation}}
\newcommand{\bea}{\begin{eqnarray}}
\newcommand{\eea}{\end{eqnarray}}
\newcommand{\bse}{\begin{subequations}}
\newcommand{\ese}{\end{subequations}}

\begin{document}

\title{Cycloidal Paths in Physics}

\author{David C. Johnston}
%\altaffiliation{johnston@ameslab.gov}
\affiliation{Ames Laboratory and Department of Physics and Astronomy, Iowa State University, Ames, Iowa 50011, USA}

\date{\today}

\begin{abstract}

A popular classroom demonstration is to draw a cycloid on a blackboard with a piece of chalk inserted through a hole at a point P with radius~$r \approx R$ from the center of a wood disk of radius~$R$ that is rolling without slipping along the chalk tray of the blackboard.  Here the parametric equations versus time are derived for the path of P from the superposition of the translational motion of the center of mass (cm) of the disk and the rotational motion of P about this cm for $r = R$ (cycloid), $r < R$ (curtate cycloid) and $r>R$ (prolate cycloid).  It is further shown that the path of P is still a cycloidal function for rolling with frictionless slipping, but where the time dependence of the sinusoidal Cartesian coordinates of the position of P is modified.  In a similar way the parametric equations versus time for the orbit with respect to a star of a moon in a circular orbit about a planet that is in a circular orbit about a star are derived, where the orbits are coplanar.  Finally, the general parametric equations versus time for the path of the magnetization vector during undamped electron-spin resonance are found, which show that cycloidal paths can occur under certain conditions.

\end{abstract}

\maketitle

\section{Introduction}

The mathematical properties of the cycloid have been studied since about 1500, and the term cycloid was evidently coined by Galileo in \mbox{$\sim 1600$~\cite{Whitman1943, Roidt2011}}.  As part of this early work, the tautochrone problem was solved, which showed that an inverted cycloid is the shape for which a particle under the influence of gravity in contact with the frictionless cycloid surface takes the same amount of time to reach the bottom of the cycloid, irrespective of its starting position~\cite{McKinley1979}.  A curtate cycloid is one in which the cusps at the bottom of the cycloid become rounded, and in a prolate cycloid the cusps are replaced by loops so that the path crosses itself periodically.  A detailed study of the cycloid and various cycloidal paths was published in 1878 and is still available~\cite{Proctor1878}.

Cycoidal paths often occur in practical physics problems. For example, the path of a charged particle starting from rest in uniform static crossed electric and magnetic fields is a cycloid~\cite{Bruce1997}.  An introductory physics classroom demonstration is to affix a light source to the edge of a circular disk and roll the disk without slipping on a flat surface \cite{Miller1959}, where the light source then traces out a cycloid.  Alternatively, one can put a piece of chalk through a hole near the edge of a circular wooden disk and roll the disk along the chalk tray of the blackboard without slipping, and the chalk then traces out a cycloid on the blackboard.  In Sec.~\ref{Sec:RollingNotSlipping} we derive the parametric equations of the Cartesian coordinates of point P versus time for such cycloidal paths in terms of the variables of this demonstration, derived from the superposition of the constant-velocity translational motion of the center of mass (cm) of the disk of radius~$R$ and the rotational motion of a point P affixed to the disk that is a distance~$r$ from the cm.  This procedure enables curtate and prolate cycloid paths to be generated in addition to the above-mentioned cycloid path, and nicely demonstrates that the motion of a point P in or on a rigid body moving through space is a superposition of the translational motion of the center-of-mass (cm) of the body and the rotational motion of the point P about the cm.   In Sec.~\ref{Sec:RollingWslipping} we show that if the disk rolls with frictionless slipping, the path of the point P is still a cycloidal function, but with a modified time dependence of the sinusoidal Cartesian coordinates of the position of P\@.

A nice simulation of the motion of a moon orbiting a planet that in turn orbits a star, where the orbits are coplanar, is available that shows cycloidal orbits  of the moon about the star~\cite{PhET}.  In Sec.~\ref{Sec:RollingWslipping} the parametric equations for the Cartesian coordinates of the moon are determined, again using the principle that the motion of the moon is the superposition of the rotational motion of the moon with respect to the planet and the translational motion of the planet in its orbit about the star.  The solution yields  epi-cycloid, epi-curtate-cycloid, or epi-prolate-cycloid paths depending on the ratio of the moon-planet distance to the planet-star distance and the relative rotational periods of moon about the planet and the planet about the star.  We show from the known parameters for our Moon-Earth-Sun system, with orbits that approximate the above idealization, that the orbit of the Moon about the Sun is a curtate cycloid with very small oscillation amplitude relative to the Earth-Sun distance.

In both nuclear magnetic resonance (NMR) \cite{Abragam1961, Slichter1963} and electron spin resonance (ESR) \cite{Pake1973, Taylor1975, Barnes1981, Poole1983} experiments, one measures how the atomic electron (ESR) or nuclear (NMR) magnetization~{\bf M} (average magnetic moment per unit volume) precesses in the presence of an applied static magnetic field {\bf H} and a continuous-wave circularly-polarized microwave magnetic field~${\bf H}_1$ that is aligned perpendicular to {\bf H}\@.  In the absence of damping, the head of the {\bf M} vector follows a time-dependent path that is determined by both {\bf H} and~${\bf H}_1$.  This path has been qualitatively described and shown in several figures~\cite{Pake1973}, but parametric equations for the path versus time have not been reported before to our knowledge.  In Sec.~\ref{Sec:ESR} the parametric equations for the path versus time are derived for undamped ESR in the general case of relevant parameters.  For $H_1\ll H$ and at the resonant frequency, the path is a helical path on the surface of a sphere as shown in Sec.~\ref{Sec:ESR}. However, when $H_1$ is a significant fraction of~$H$, a cycloidal path is found, the nature of which depends on the parameters of the electron-spin resonance.  A summary is given in Sec.~\ref{Sec:Summary}.

\section{Results: Cycloidal Paths}

\subsection{\label{Sec:RollingNotSlipping} Rolling without Slipping}

\begin{figure}
\includegraphics [width=2.5in]{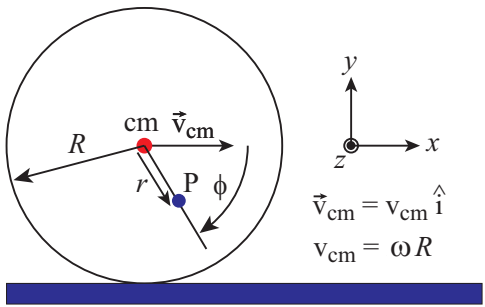}
\caption {Experimental configuration of a disk of radius~$R$ rolling towards the right on a horizontal surface without slipping.  The speed of the center of mass (cm) of the disk is $v_{\rm cm}=\omega R$ and hence the velocity of the cm is ${\bf v}_{\rm cm}=\omega R \, \hat{\bf i}$.  The point P is a distance~$r$ from the cm at a negative instantaneous polar angle $\phi$ measured clockwise from the positive $x$~axis.}
\label{Fig:Cycloidal_functions}
\end{figure}

The experimental configuration is shown in Fig.~\ref{Fig:Cycloidal_functions}.  The disk is rolling without slipping towards the right.  The azimuthal angle $\phi$ of the point P is measured with respect to the positive $x$~axis and is negative because P is rotating clockwise instead of counterclockwise for which $\phi$ would be positive.  Because the disk is rolling without slipping, the constant speed of the center of mass (cm) with respect to the stationary surface on which the disk rolls is
\be
v_{\rm cm} = \omega R,
\ee
where $\omega = |d\phi/dt|$ is the angular speed of P with respect to the cm.  The position ${\bf r}_{\rm cm}$ of the cm is
\bse
\label{Eqs:r}
\be
{\bf r}_{\rm cm} = \omega R t\,\hat{\bf i},
\label{Eq:rcm}
\ee
where $t$ is the time.  The rotational motion of P with respect to the cm is described by
\be
{\bf r}_{\rm P,\,rot} = r\left[\cos(\omega_zt)\,\hat{\bf i} + \sin(\omega_zt)\,\hat{\bf j}\right].
\ee
Since $\omega_z = -\omega$, one obtains
\be
{\bf r}_{\rm rot} = r\left[\cos(\omega t)\,\hat{\bf i} - \sin(\omega t)\,\hat{\bf j}\right].
\label{Eq:rProt}
\ee
\ese

The position ${\bf r}$ of point~P is the superposition of the center of mass position and the rotational position in Eqs.~(\ref{Eq:rcm}) and~(\ref{Eq:rProt}), respectively.  The Cartesian components of ${\bf r}$ are therefore
\bse
\bea
x &=& \omega R t + r \cos(\omega t),\\
y &=&- r \sin(\omega t).
\eea
\ese
Dividing both sides of each of these equations by $R$ gives dimensionless components
\bse
\label{Eqs:Cycloidal}
\bea
\frac{x}{R} &=& \omega t + \frac{r}{R} \cos(\omega t),\\
\frac{y}{R} &=&- \frac{r}{R} \sin(\omega t).
\eea
\ese
These are the parametric equations for the reduced Cartesian components in terms of the implicit parameter $\omega t$.  The amplitudes of the sinusoidal components of $x/R$ and~$y/R$ have the same value $r/R$.

The parametric equations for the cycloid are conventionally written~\cite{Roidt2011}
\bse
\bea
\frac{x}{R} &=& \theta - \sin\theta,\\
\frac{y}{R} &=& 1-\cos\theta.
\eea
\ese
One can obtain these equations from Eqs.~(\ref{Eqs:Cycloidal}) by the substitutions $\omega t\to \theta + \pi/2$, $r/R \to 1$, and a $y$-axis offset $y/R\to y/R +1$.

\begin{figure}
\includegraphics [width=3.3in]{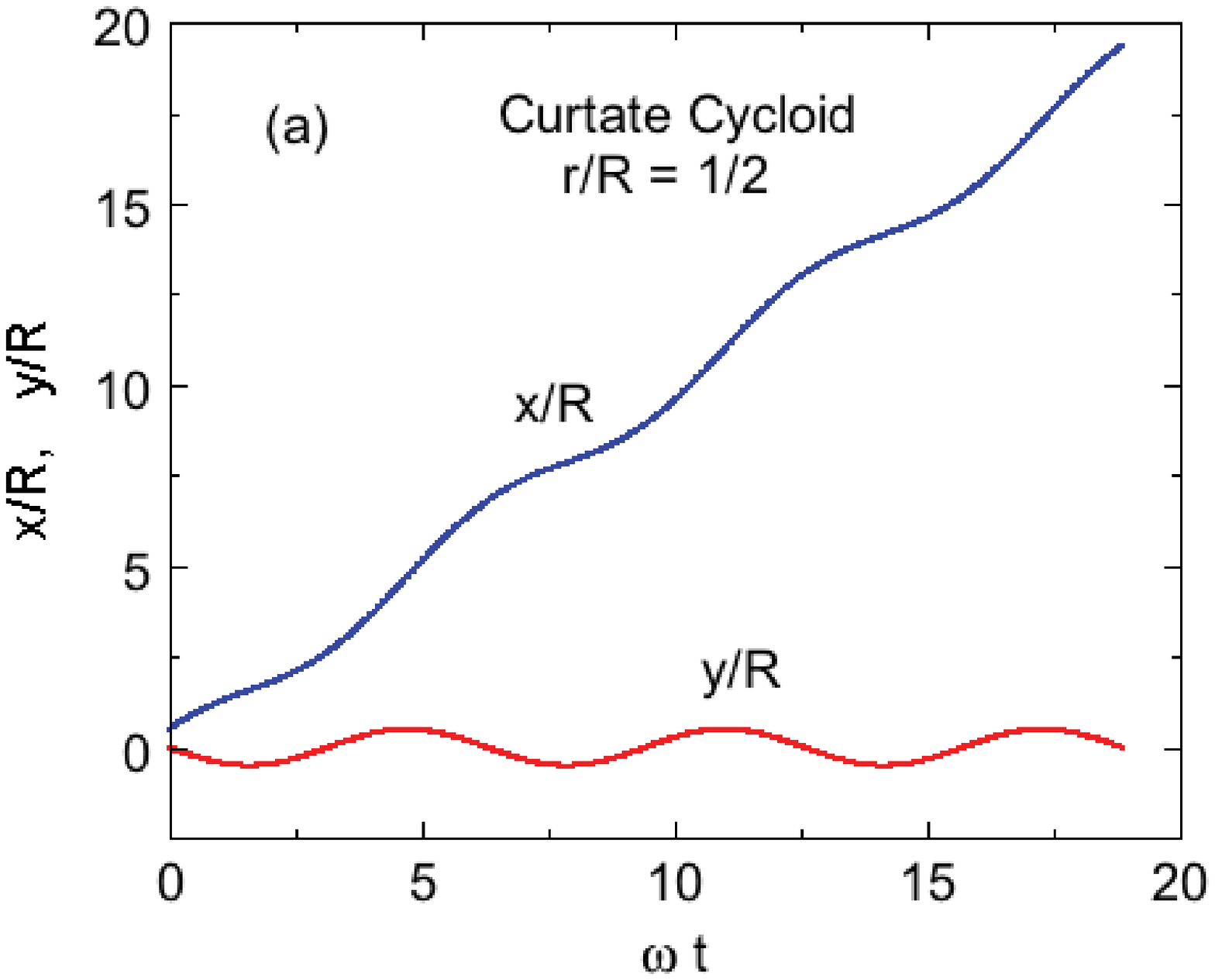}
\includegraphics [width=3.3in]{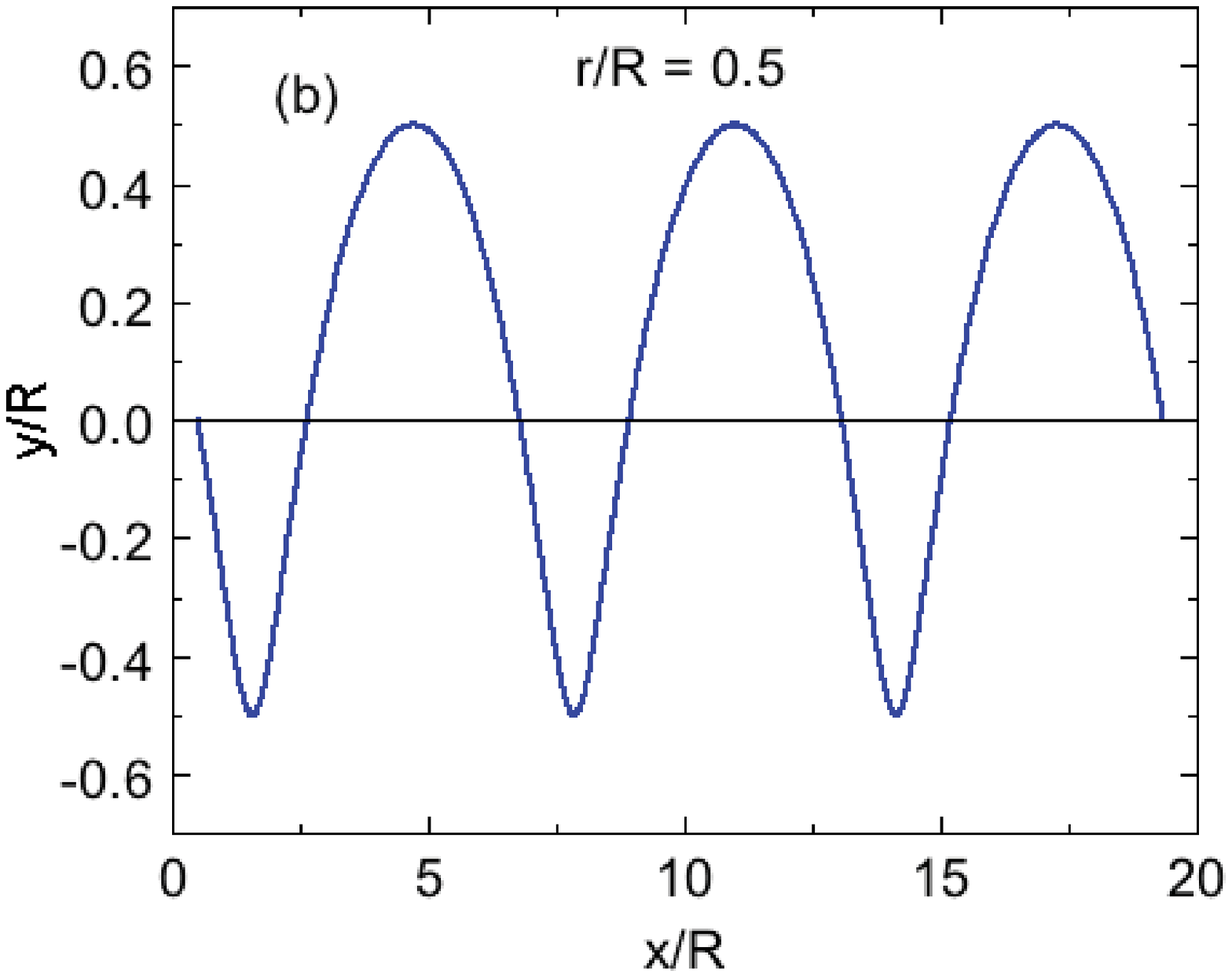}
\caption {(a) Cartesian components $x/R$ and $y/R$ of point P versus reduced time $\omega t$ in radians for $r/R=0.5$. (b)~$y/R$ versus $x/R$ with $\omega t$ as an implicit parameter where this path of point P is a curtate cycloid.  These plots were obtained using Eqs.~(\ref{Eqs:Cycloidal}).}
\label{Fig:cycloid_rONR0.5}
\end{figure}

\begin{figure}
\includegraphics [width=3.35in]{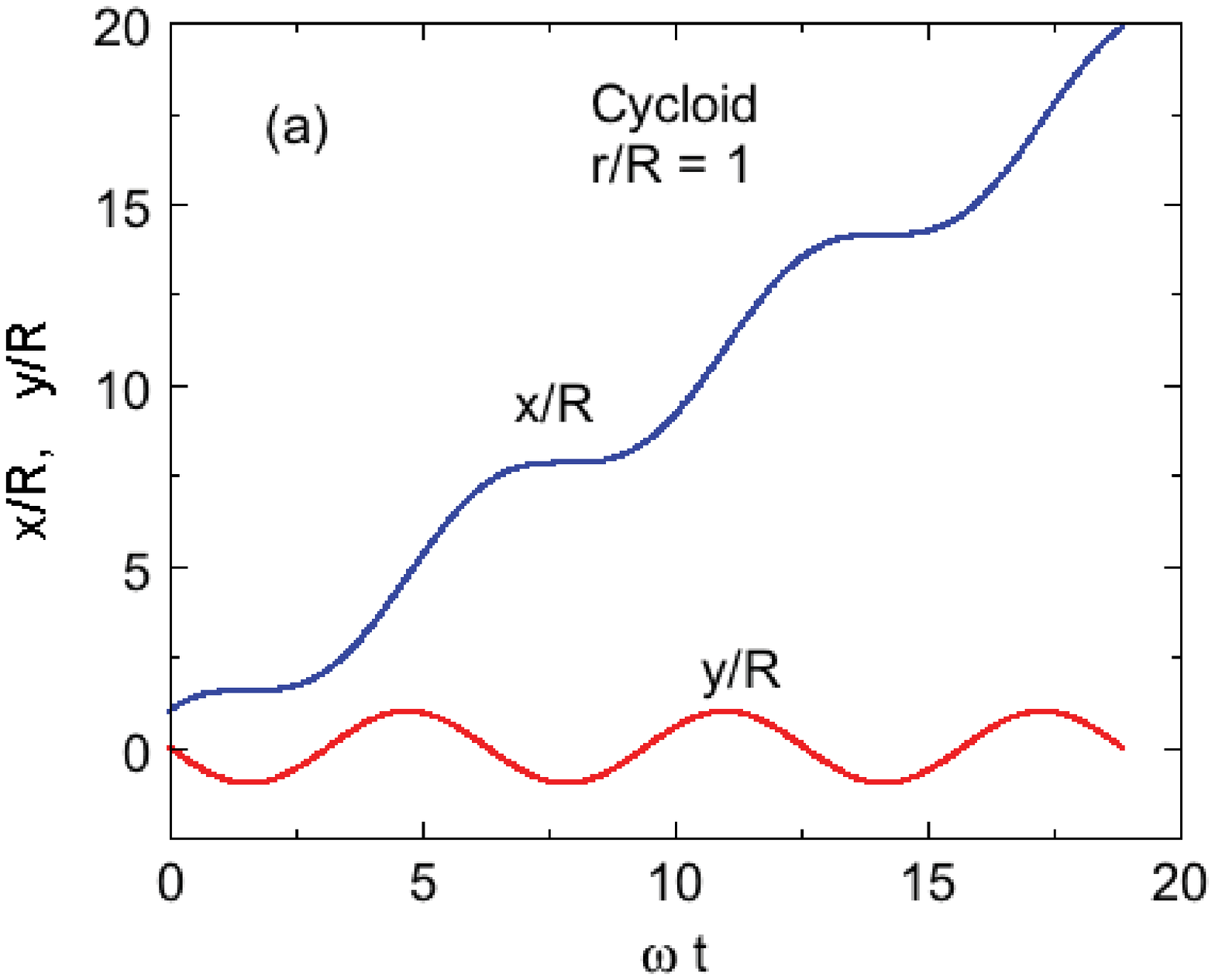}
\includegraphics [width=3.35in]{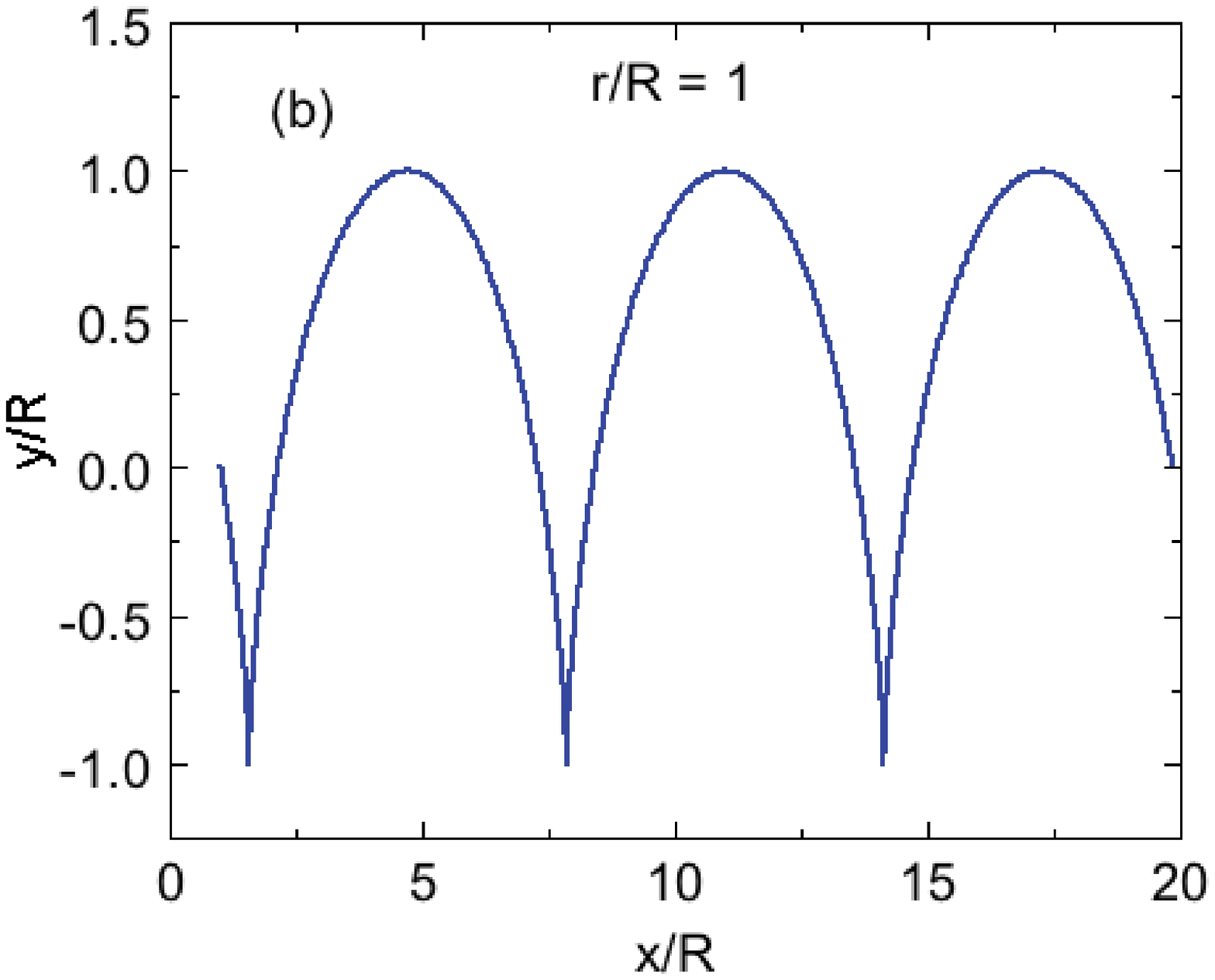}
\caption {(a) Cartesian components $x/R$ and $y/R$ of point P versus reduced time $\omega t$ in radians for $r/R=1$. (b)~$y/R$ versus $x/R$ with $\omega t$ as an implicit parameter where this path of point P is a cycloid.    These plots were obtained using Eqs.~(\ref{Eqs:Cycloidal}).}
\label{Fig:cycloid_rONR1.0}
\end{figure}

\begin{figure}
\includegraphics [width=3.3in]{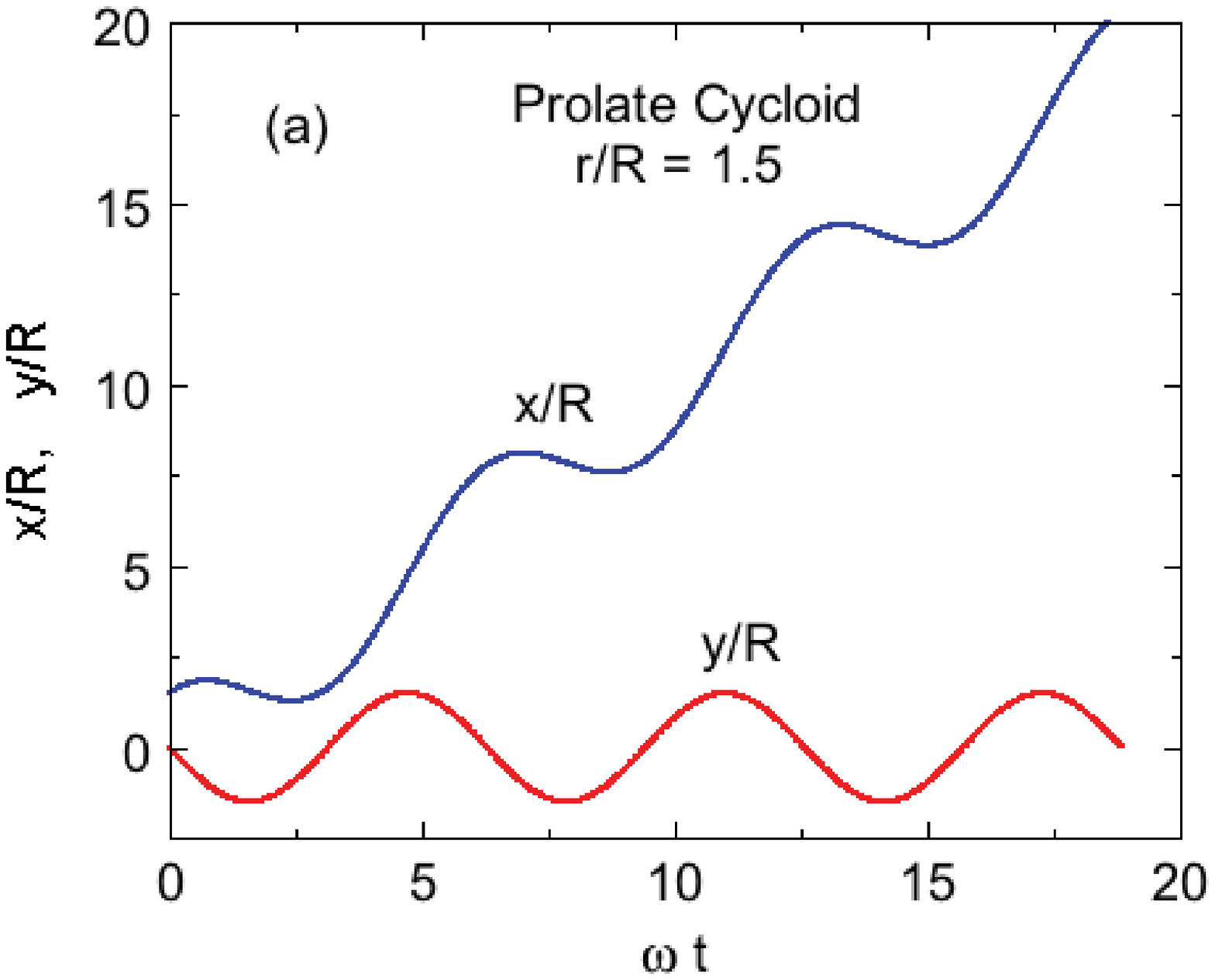}
\includegraphics [width=3.3in]{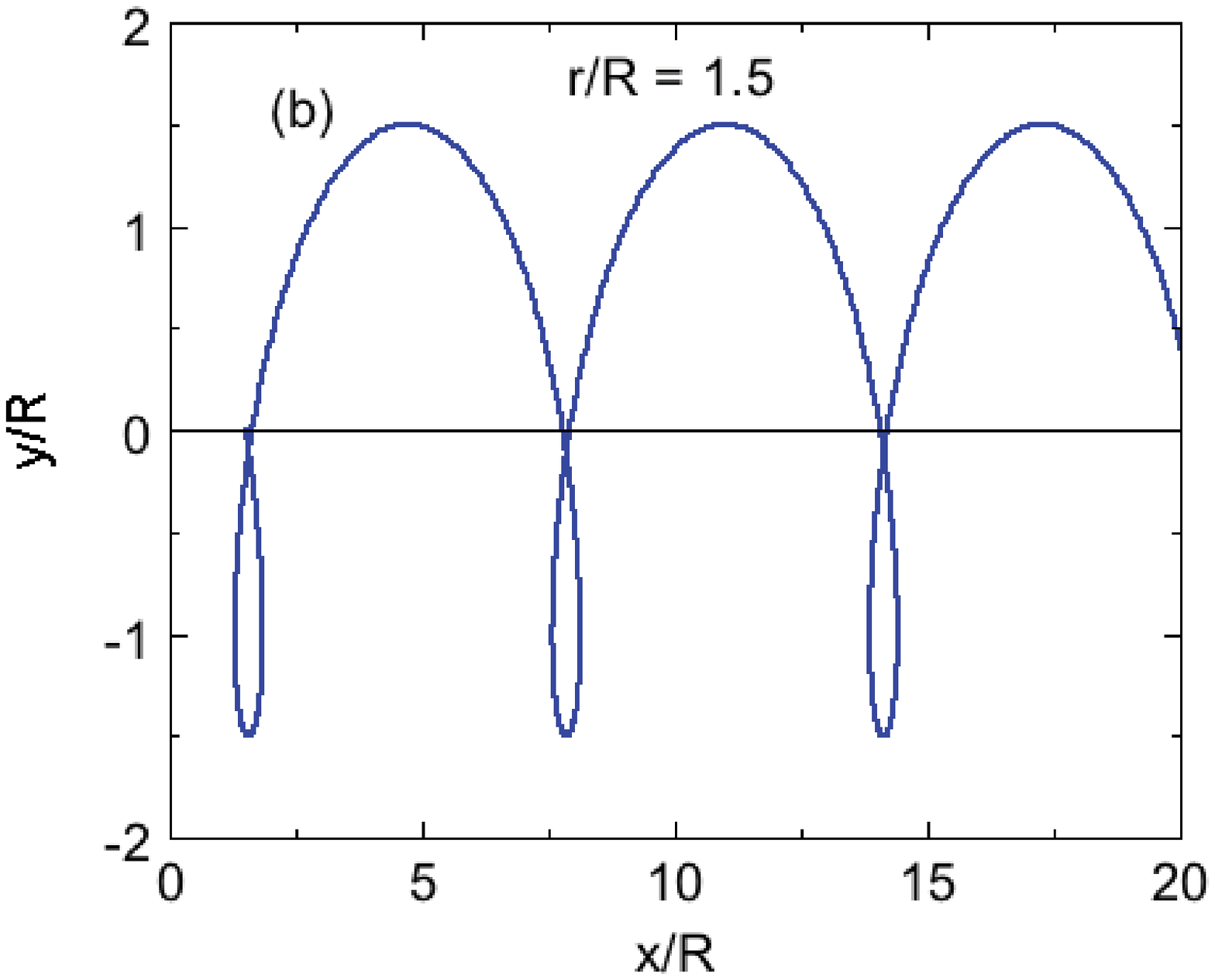}
\caption {(a) Cartesian components $x/R$ and $y/R$ of point P versus reduced time $\omega t$ in radians for $r/R=1.5$. (b)~$y/R$ versus $x/R$ with $\omega t$ as an implicit parameter where this path of point P is a prolate cycloid.    These plots were obtained using Eqs.~(\ref{Eqs:Cycloidal}).}
\label{Fig:cycloid_rONR1.5}
\end{figure}

Shown in Fig.~\ref{Fig:cycloid_rONR0.5}(a) are plots of $x/R$ and~$y/R$ versus $\omega t$ for a curtate cycloid with $r/R=0.5$ using Eqs.~(\ref{Eqs:Cycloidal}).  Figure~\ref{Fig:cycloid_rONR0.5}(b) shows the parametric plot of $y/R$ versus~$x/R$ with $\omega t$ as the implicit parameter.  A curtate cycloid can be traced on the blackboard by drilling a hole in the wooden disk at radius $r<R$ in which to insert a piece of chalk and then rolling the disk along the chalk tray as in the above procedure to generate a chalk trace of a cycloid.  Corresponding plots for the cycloid with $r/R = 1$ are shown in Fig.~\ref{Fig:cycloid_rONR1.0}.  The derivative $dy/dx$ of the cycloid curve in Fig.~\ref{Fig:cycloid_rONR1.0}(b) is discontinuous at the minima, whereas for the curtate cycloid in Fig.~\ref{Fig:cycloid_rONR0.5}(b) the minima are rounded.  Because Figs.~\ref{Fig:cycloid_rONR0.5}(a) and~\ref{Fig:cycloid_rONR1.0}(a) are so similar, one might not anticipate the significant difference between Figs.~\ref{Fig:cycloid_rONR0.5}(b) and~\ref{Fig:cycloid_rONR1.0}(b).  At sufficiently small values of $r/R$ the $y$ versus $x$ curves become nearly sinusoidal.    Figure~\ref{Fig:cycloid_rONR1.5} shows corresonding plots for a prolate cycloid with $r/R = 1.5$.  Here loops appear in the $y/R$ versus $x/R$ plot in Fig.~\ref{Fig:cycloid_rONR1.5}(b).

\subsection{\label{Sec:RollingWslipping} Rolling with Frictionless Slipping}

Here we consider a disk of radius~$R$ that is rotating at angular speed $\omega^\prime\neq v_{\rm cm}/R$ and thus rolling with slipping without friction on a surface.  Here we ask the same question as in this last section: what is the path through space of a point P that is fixed on the disk a distance~$r$ from its center? We initially assume that the angular velocity of the disk is in the same direction as in the previous section for rolling without slipping, but the following results are easily generalized to the case where the angular velocity of the disk with slipping is in the opposite direction of the case without slipping by simply changing the sign of the parameter~$\alpha$ introduced below from positive to negative.

Referring again to Fig.~\ref{Fig:Cycloidal_functions}, here we write $v_{\rm cm} = \omega R$ where $\omega$ is the angular speed of the disk if it were rolling without slipping.  Therefore one again has
\bse
\be
{\bf r}_{\rm cm} = \omega R t\,\hat{\bf i}
\label{Eq:rcm2}
\ee
However, the rotational motion of P with respect to the cm is now described by
\bea
{\bf r}_{\rm rot} &=& r\left[\cos(\omega_zt)\,\hat{\bf i} + \sin(\omega_zt)\,\hat{\bf j}\right]\\
&=& r\left[\cos(\omega^\prime t)\,\hat{\bf i} - \sin(\omega^\prime t)\,\hat{\bf j}\right],
\label{Eq:rProt2}
\eea
\ese
where $\omega^\prime$ is the angular speed of the disk which satisfies $\omega^\prime\neq\omega$ for rolling with slipping.  The reduced $x$ and~$y$ components of {\bf r} are now
\bse
\label{Eqs:xyp}
\bea
\frac{x}{R} &=& \omega t + \frac{r}{R} \cos(\omega^\prime t),\\
\frac{y}{R} &=&- \frac{r}{R} \sin(\omega^\prime t).
\eea
\ese
We write the relationship between $\omega^\prime$ and~$\omega$ as
\be
\omega^\prime=\alpha\omega,
\ee
where $\alpha$ is a dimensionless constant. Then Eqs.~(\ref{Eqs:xyp}) become
\bea
\frac{x}{R} &=& \omega t + \frac{r}{R} \cos(\alpha\omega t), \label{Eqs:xyGen}\\
\frac{y}{R} &=& - \frac{r}{R} \sin(\alpha\omega t). \nonumber
\eea

\begin{figure}
\includegraphics [width=3.3in]{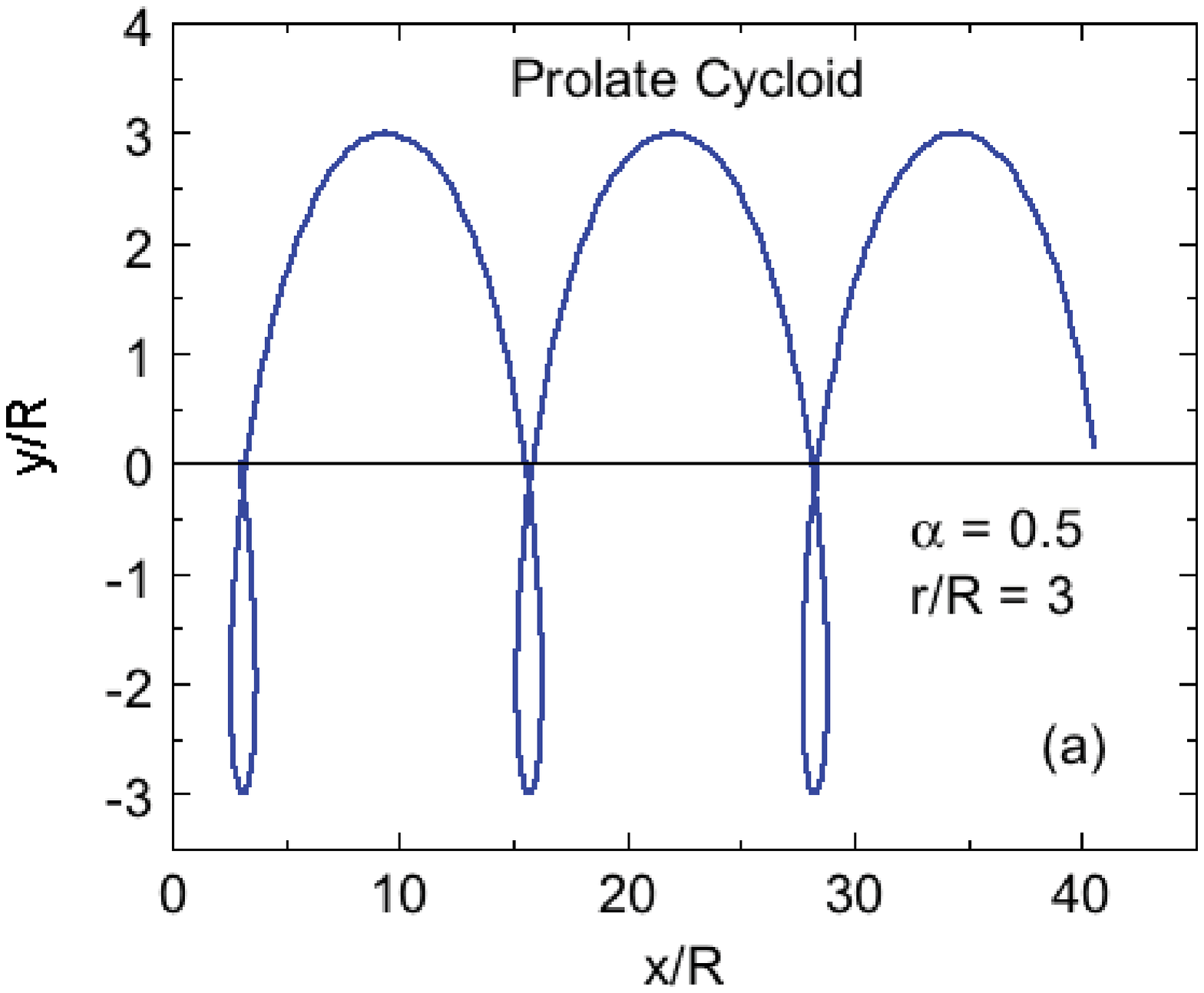}
\includegraphics [width=3.3in]{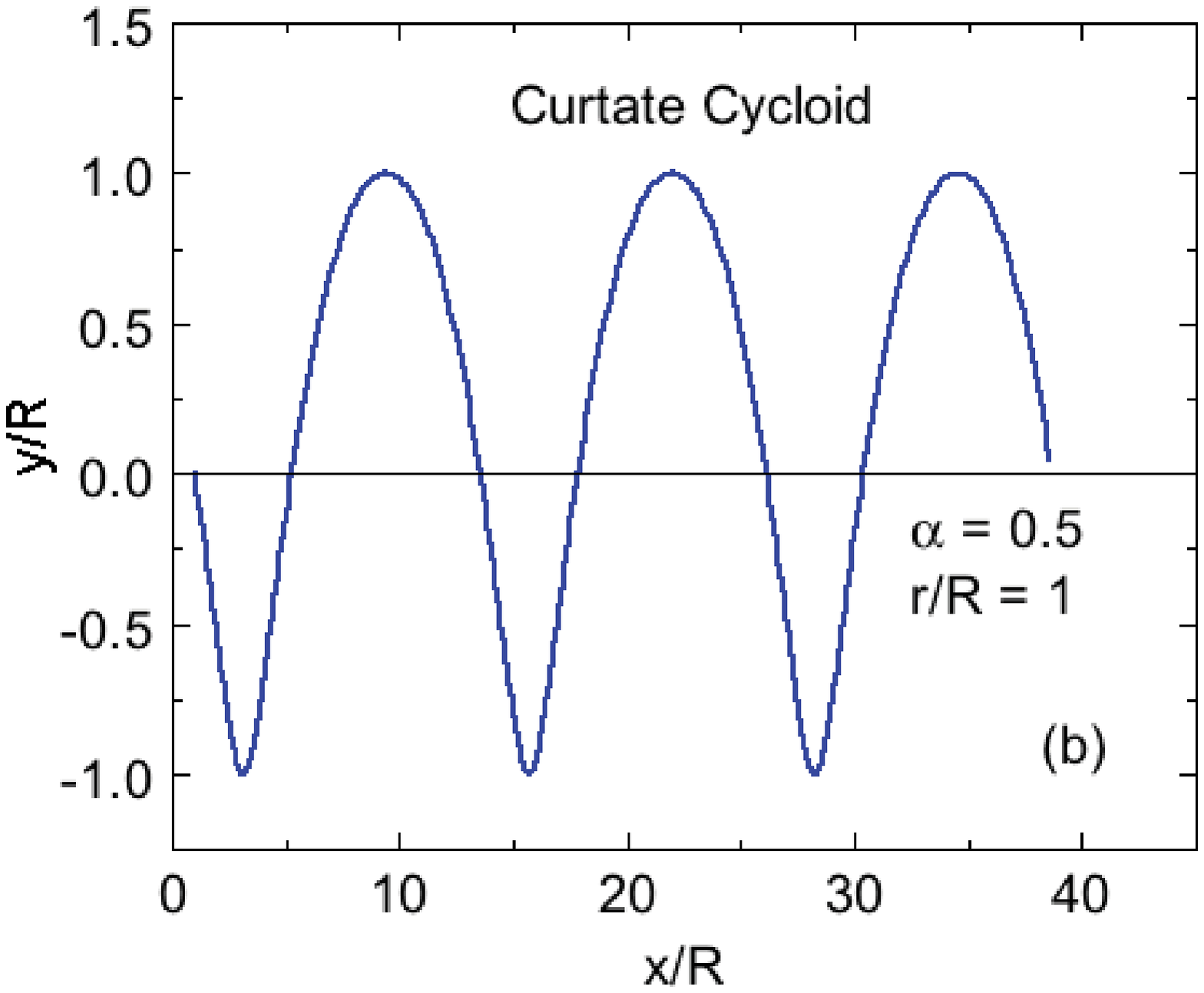}
\includegraphics [width=3.3in]{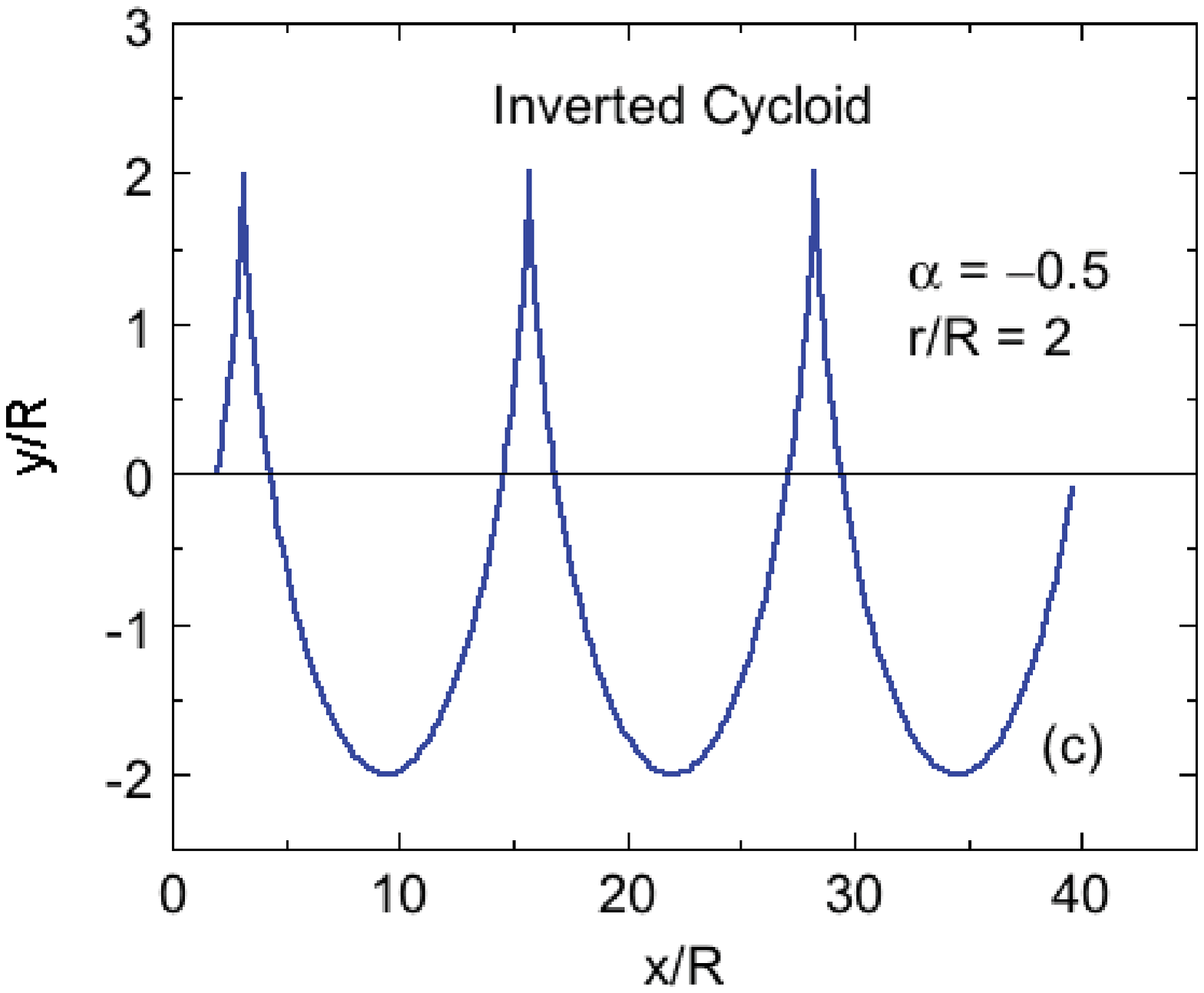}
\caption {Cartesian component $y/R$ versus $x/R$ of the path of point P with $T = \alpha\omega t$ as an implicit parameter for (a)~$\alpha = 1/2$ and $r/R = 3$ (prolate cycloid), (b)~$\alpha = 1/2$ and $r/R = 1$ (curtate cycloid), and (c)~$\alpha = -1/2$ and $r/R = 2$ (inverted cycloid) calculated using Eqs.~(\ref{Eqs:xyGen}).}
\label{Fig:Cycloid_arOnR1.5_a0.5}
\end{figure}

Thus the motion of point P when the disk is rolling at constant angular speed with frictionless slipping is a cycloidal function, but where the time dependence of the sinusoidal parts of $x$ and~$y$ are changed in the same way compared with the case of rolling without slipping.  In particular, the path is a curtate cycloid if $|\alpha| r/R < 1$, a cycloid if $|\alpha| r/R = 1$, and a prolate cycloid if $|\alpha| r/R > 1$.   For the case where the disk is rotating in the opposite direction while slipping compared to the case of rolling without slipping, one replaces the positive $\alpha$ in Eqs.~(\ref{Eqs:xyGen}) by $-\alpha$ and the path becomes inverted.  Example plots of  Eqs.~(\ref{Eqs:xyGen}) are shown in Fig.~\ref{Fig:Cycloid_arOnR1.5_a0.5} for $\alpha = 1/2$ and $r/R = 3$ (prolate cycloid),  $\alpha = 1/2$ and $r/R = 1$ (curtate cycloid), and  $\alpha = -1/2$ and $r/R = 2$ (inverted cycloid).

\subsection{Path of a Moon with respect to a Star while Orbiting a Planet that is Orbiting the Star}

Here we consider the coplanar orbits of a moon orbiting a planet while the planet orbits a star that is stationary with respect to the distant stars, where the moon, planet and star are spherically symmetric, the two orbits are circular and lie in the $xy$~plane, and the moon and planet are both moving counterclockwise in their orbits when viewed from the positive $z$~axis.  The Moon orbiting the Earth that orbits the Sun approximately satisfies these conditions and this case will be discussed below.

We first define the following abbreviations for this section:\\
$r = $ moon to planet distance (center to center)\\
$R = $ planet to star distance (center to center)\\
$T_{\rm M} =$ period of the moon's orbit about the planet\\
$T_{\rm P} =$ period of the planet's orbit about the star\\
$\omega_{\rm M} = 2\pi/T_{\rm M} = $ angular speed of the moon with respect\\
\indent\indent to the planet\\
$\omega_{\rm P} = 2\pi/T_{\rm P} = $ angular speed of the planet with respect\\
\indent\indent  to the star\\
$(x,y) = $ Cartesian coordinates of the moon's center with\\
\indent\indent respect to the star\\
$(x_{\rm P},y_{\rm P}) = $ Cartesian coordinates of the planet's center\\
\indent\indent with respect to the star\\
One expects $r\ll R$, $T_{\rm M}\ll T_{\rm P}$, $\omega_{\rm M}\gg \omega_{\rm P}$.

Thus we have
\bse
\label{Eqs:xy}
\bea
x_{\rm P} &=& R\cos(\omega_{\rm P}t),\\
y_{\rm P} &=& R\sin(\omega_{\rm P}t),\\
x &=& r \cos(\omega_{\rm M}t) + R\cos(\omega_{\rm P}t),\label{Eq:xMoon}\\
y &=& r \sin(\omega_{\rm M}t) + R\sin(\omega_{\rm P}t).\label{Eq:yMoon}
\eea
\ese
We define the dimensionless parameter \mbox{$\alpha \equiv \omega_{\rm P}/\omega_{\rm M}\ll 1$} and from Eqs.~(\ref{Eq:xMoon}) and~(\ref{Eq:yMoon}) obtain 
\bse
\bea
x &=& r \cos(\omega_{\rm M}t) + R\cos(\alpha\omega_{\rm M}t),\\
y &=& r \sin(\omega_{\rm M}t) + R\sin(\alpha\omega_{\rm M}t).
\eea
\ese
Using the generic expression $\omega = 2\pi/T$ one obtains
\bse
\label{Eqs:xy2}
\bea
x &=& r \cos(2\pi t/T_{\rm M}) + R\cos(2\pi\alpha t/T_{\rm M}),\\
y &=& r \sin(2\pi t/T_{\rm M}) + R\sin(2\pi\alpha t/T_{\rm M}).
\eea
\ese
Finally, introducing the dimensionless reduced time
\be
T \equiv t/T_{\rm M}
\ee
and dividing both sides Eqs.~(\ref{Eqs:xy2}) by $R$, one obtains the dimensionless parametric equations for the path of the moon with respect to the star versus reduced time~$T$ as
\bse
\label{Eqs:Redxy}
\bea
\frac{x}{R} &=&  \frac{r}{R}\cos(2\pi T) + \cos(2\pi\alpha T),\\
\frac{y}{R} &=&  \frac{r}{R}\sin(2\pi T) + \sin(2\pi\alpha T).
\eea
\ese

\begin{figure}
\includegraphics [width=2.8in]{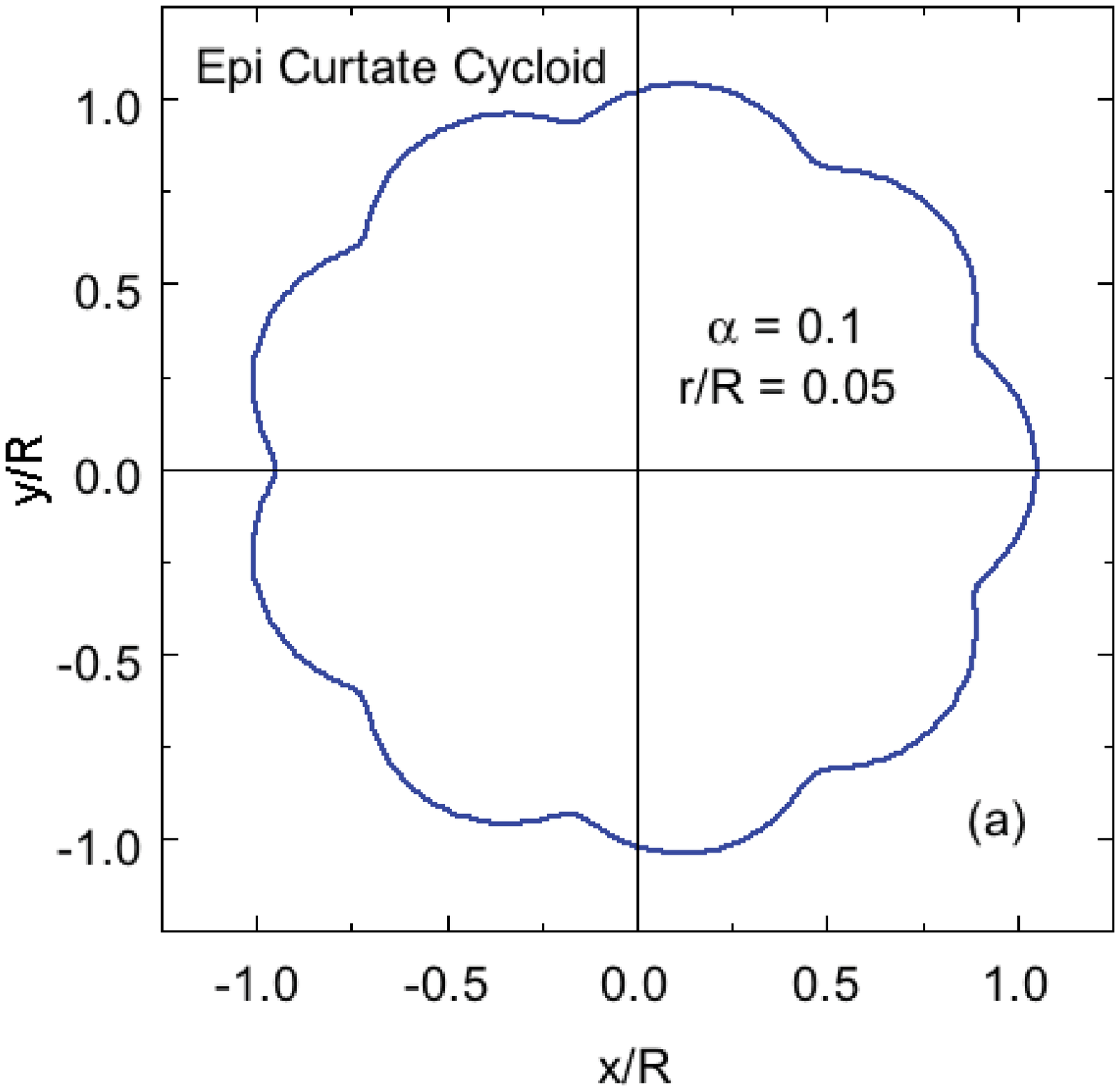}
\includegraphics [width=2.8in]{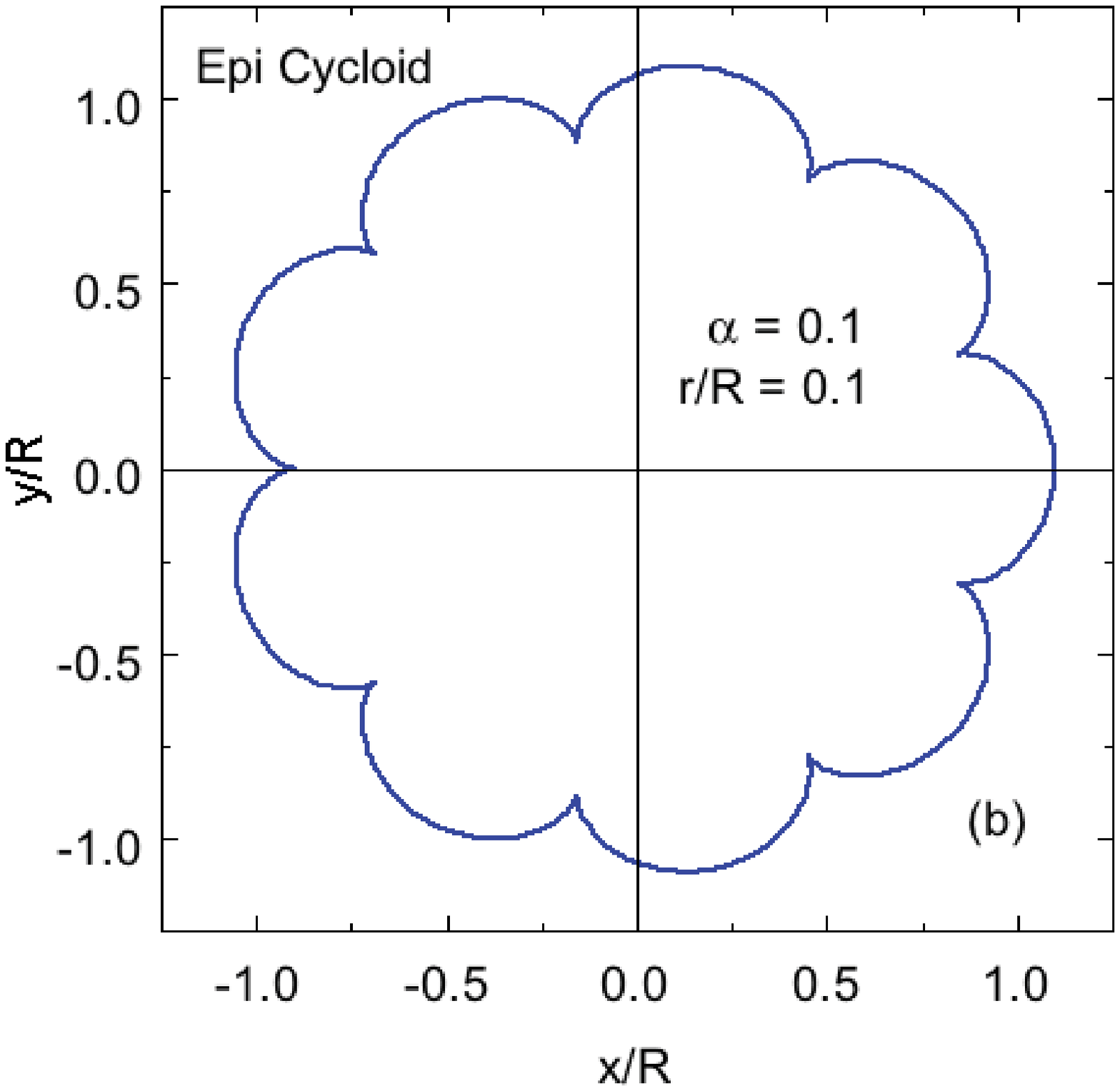}
\includegraphics [width=2.8in]{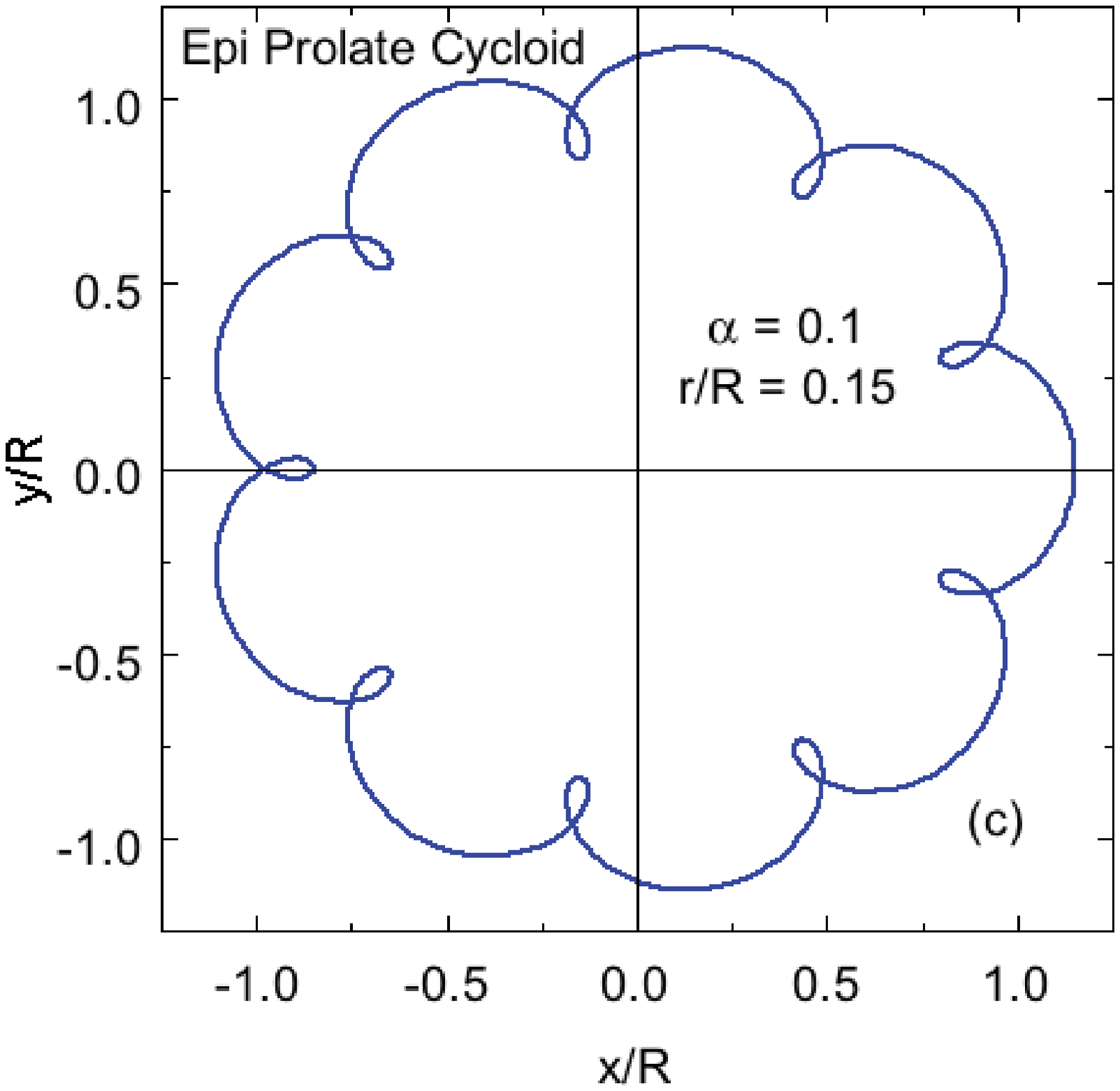}
\caption {Parametric plots of $y/R$ versus $x/R$ for cycloidal paths of a moon orbiting a planet that orbits a star with $\alpha = 0.1$ and (a)~$r/R = 0.05$ (epi-curtate cycloid), (b)~$r/R = 0.10$ (epi-cycloid), and (c)~$r/R = 0.15$ (epi-prolate cycloid),  calculated using Eqs.~(\ref{Eqs:Redxy}).}
\label{Fig:ycloid_alph0.1_ronR0.05}
\end{figure}

Parametric plots of $y/R$ versus $x/R$ using Eqs.~(\ref{Eqs:Redxy}) are shown in Fig.~\ref{Fig:ycloid_alph0.1_ronR0.05} for $\alpha=0.1$ and~$r/R = 0.05$, 0.1, and 0.15.  The latter three parameters are unrealistically large in order to clearly show the structure of the paths.  The paths are analogous to those in Figs.~\ref{Fig:cycloid_rONR0.5}--\ref{Fig:cycloid_rONR1.5}, except that  the $x$~axes in those figures are bent here into circles.  Thus for $r/R<\alpha$ one obtains an epi-curtate cycloid, for $r/R=\alpha$ an epi-cycloid, and for $r/R>\alpha$ an epi-prolate cycloid, where here the prefix epi refers to linear cycloidal motion bent into a circle.

The orbit of the Moon about the Earth and the Earth about the Sun are approximately coplanar.  When viewed from the North, the Earth rotates counter-clockwise about the Sun at a distance $R = 1.50\times 10^{11}$~m with rotation period $T_{\rm P} = 365.4$~d and angular speed $\omega_{\rm P} = 2\pi/T_{\rm P}$.  The Moon rotates counter-clockwise around the Earth at a distance $r = 3.84\times 10^8$~m with rotation period $T_{\rm M} = 27.3$~d and angular speed $\omega_{\rm M} = 2\pi/T_{\rm M}$.  Thus for the Moon orbiting the Earth, the parameter $\alpha = 27.3/365.3 \approx 0.075$, roughly the same as the value $\alpha = 0.1$ used to construct Fig.~\ref{Fig:ycloid_alph0.1_ronR0.05}.  However, the ratio $r/R \approx 0.0026$ is much smaller than the values of 0.05 to 0.15 in Fig.~\ref{Fig:ycloid_alph0.1_ronR0.05}.  Thus $r/R \ll \alpha$ and hence the Moon has an epi-curtate cycloidal path around the Sun corresponding to the linear rolling with slipping path in Fig.~\ref{Fig:Cycloid_arOnR1.5_a0.5}(b), but with a very small amplitude of oscillation (not shown) that is barely visible on the scale of the plot in Fig.~\ref{Fig:ycloid_alph0.1_ronR0.05}(a).

\subsection{\label{Sec:ESR} Paths of the Magnetization Vector in Undamped Electron-Spin Resonance}

The Bloch equations are often the starting point for analyzing experimental electron-spin resonance (ESR) data. In the absence of damping, the Bloch equations give the Cartesian components of the magnetization~{\bf M} (average magnetic moment per unit volume) that is precessing around the applied uniform, static magnetic field
\be
{\bf H} = H_0 \hat{\bf k},
\ee
as~\cite{Bloch1946}
\bse
\label{Eqs:Bloch}
\bea
\frac{dM_x}{dt} &=& -\gamma ({\bf M}\times{\bf H})_x,\\
\frac{dM_y}{dt} &=& -\gamma ({\bf M}\times{\bf H})_y ,\\
\frac{dM_z}{dt} &=& -\gamma ({\bf M}\times{\bf H})_z ,
\eea
\ese
where the negative sign prefactors arise from the negative charge on the electron appropriate for ESR, and 
$\gamma$ is the gyromagnetic ratio ($\gamma = g\mu_{\rm B}/\hbar$ for Heisenberg spins, $g$ is the spectroscopic splitting factor, $\mu_{\rm B}$ is the Bohr magneton, and $\hbar$ is Planck's constant divided by $2\pi$).  The Gaussian cgs system of units is used in this section.

In the absence of damping and additional magnetic fields, the Bloch equations yield a magnetization that precesses around ${\bf H}$ at angular frequency
\be
\omega_0=\gamma H_0
\ee
according to
\bse
\bea
M_x &=& M_0 \sin\theta\cos(\omega_0t),\\
M_y &=& M_0 \sin\theta\sin(\omega_0t),\\
M_z &=& M_0 \cos\theta,
\eea
\ese
where $M_0 = |{\bf M}|$ and $\theta$ is the constant angle that {\bf M} makes with the $z$~axis during the precession.

For ESR experiments, an additional circularly-polarized microwave magnetic field ${\bf H}_1$ with angular frequency $\omega$ is applied that rotates in the $xy$~plane about the $z$~axis in the same direction that {\bf M} is precessing in the absence of ${\bf H}_1$, given by
\be
{\bf H}_1 = -H_1[\cos(\omega t)\hat{\bf i} + \sin(\omega t)\hat{\bf j}] \label{Eq:RC},
\ee
where the negative-sign prefactor is due to the negative charge on the electron that applies to ESR as in Eqs.~(\ref
{Eqs:Bloch}).  Thus ${\bf H}_1$ is always antiparallel to the projection of {\bf M} onto the $xy$~plane.

\begin{figure}
\includegraphics [width=2in]{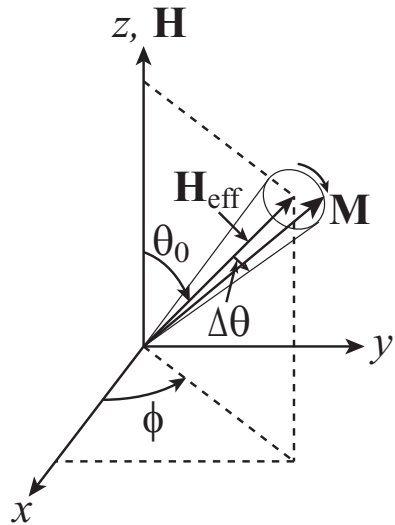}
\caption {Geometry of the precession of the magnetization {\bf M} in the presence of an applied static field {\bf H} and a microwave magnetic field ${\bf H}_1$.  {\bf M} precesses clockwise about the effective field ${\bf H}_{\rm eff}$, forming the surface of a cone  with cone half-angle $\Delta \theta$, while ${\bf H}_{\rm eff}$ precesses counter-clockwise about the applied field {\bf H} with azimuthal angle $\phi = \omega t$ on the surface of a cone with cone angle $\theta_0$ given by Eq.~(\ref{conehalfangle}).  After Ref.~\cite{Pake1973b}.  }
\label{Fig:Undamped_Precession}
\end{figure}

Qualitatively, {\bf M} precesses around an effective magnetic field ${\bf H}_{\rm eff}$ at angular frequency $\omega_{\rm eff}$ while ${\bf H}_{\rm eff}$ precesses around the applied field~{\bf H} at the angular frequency~$\omega$ as shown in Fig.~\ref{Fig:Undamped_Precession}, where~\cite{Pake1973b}
\bse
\label{Eqs:Hsomegas}
\bea
{\bf H}_{\rm eff} &=& -H_1[\cos(\omega t)\,\hat{\bf i} + \sin(\omega t)\,\hat{\bf j}]+ \left(H_0 - \frac{\omega}{\gamma}\right)\,\hat{\bf k},\hspace{0.3in} \\
H_{\rm eff} &=& \left|{\bf H}_{\rm eff}\right| = \sqrt{H_1^2+\left(H_0-\frac{\omega}{\gamma}\right)^2},\\
\omega_{\rm eff} &=& \gamma H_{\rm eff} = \sqrt{\omega_1^2+(\omega_0 - \omega)^2}\\
\omega_1 &\equiv& \gamma H_1,\\
\phi &=& \omega t,\\
\theta_0 &=& {\rm arctan}\left(\frac{\omega_0 - \omega}{\omega_1}\right) \quad (0\leq \theta_0 \leq \pi/2),\label{conehalfangle}\\
\theta &=& \theta_0 + \Delta\theta \cos\left(\omega_{\rm eff} t\right),
\eea
\ese
and the cone half-angle $\Delta\theta$ with $0<\Delta\theta\leq \frac{\pi}{2} - \theta_0$ is an adjustable parameter.  Here we derive an expression for the  path that the head of the magnetization vector {\bf M} follows when both {\bf H} and ${\bf H}_1$ are present. 

For subsequent calculations and plots, we normalize all angular frequencies by $\omega_0$, the time by $1/\omega_0$, and the magnetization magnitude $M$ by $M_0$, yielding the dimensionless reduced parameters
\bse
\bea
\bar{M} &=& 1,\\
T &=& \omega_0 t,\\
\bar{\omega}_0 &=& 1,\\
\bar{\omega} &=& \omega/\omega_0,\\
\bar{\omega}_1 &=& \omega_1/\omega_0,\\
\bar{\omega}_{\rm eff} &=& \omega_{\rm eff}/\omega_0.
\eea
\ese
From Eqs.~(\ref{Eqs:Hsomegas}) one then obtains
\bse
\bea
\bar{\omega}_{\rm eff} &=& \sqrt{\bar{\omega}_1^2 + \left(1-\bar{\omega}^2\right)}, \label{Eq:omegaeff}\\
\theta &=& \theta_0 + \Delta\theta\cos(\bar{\omega}_{\rm eff}T),\\
\phi &=& \bar{\omega} T,\\
\theta_0 &=& {\rm arctan}\left(\frac{\bar{\omega}_1}{1 - \bar{\omega}}\right).
\eea
\ese

\begin{figure}
\includegraphics [width=1.75in]{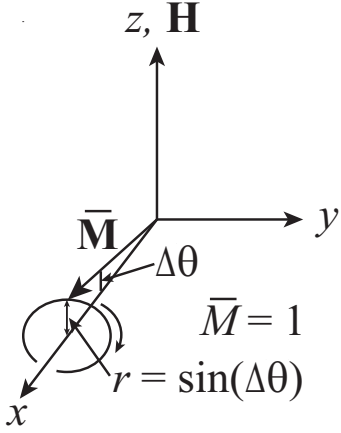}
\caption {First step of generating the path of the precessing magnetization~$M$ in Fig.~\ref{Fig:Undamped_Precession}.}
\label{Fig:Undamped_Precession_1}
\end{figure}

To generate the path of the head of {\bf M} versus time according to Fig.~\ref{Fig:Undamped_Precession}, we first consider the configuration in Fig.~\ref{Fig:Undamped_Precession_1} where the initial position ${\bf M}^{\rm A}$ of $\bar{\bf M}$ at time $t=0$ is 
\bea
\bar{\bf M}^{\rm A}_x &=& \cos\Delta\theta,\nonumber\\
\bar{\bf M}^{\rm A}_y &=& 0,\\
\bar{\bf M}^{\rm A}_z &=& \sin\Delta\theta.\nonumber
\eea
Rotating ${\bf M}^{\rm A}$ clockwise about the $x$~axis by the negative angle $-\bar{\omega}_{\rm eff}T$ gives the precession of {\bf M} about the $x$~axis as
\bea
\bar{\bf M}^{\rm B}_x &=& \cos\Delta\theta,\nonumber\\
\bar{\bf M}^{\rm B}_y &=&  \sin\Delta\theta\sin(\bar{\omega}_{\rm eff}T),\\
\bar{\bf M}^{\rm B}_z &=& \sin\Delta\theta\cos(\bar{\omega}_{\rm eff}T).\nonumber
\eea
Next we rotate ${\bf M}^{\rm B}$ clockwise about the $y$~axis by a negative angle $-\left(\frac{\pi}{2}-\theta_0\right)$ so that ${\bf H}_{\rm eff}$ is at an angle of $\theta_0$ with respect to the $z$~axis according to Fig.~\ref{Fig:Undamped_Precession}, yielding
\bea
\bar{\bf M}^{\rm C}_x &=& -\sin\Delta\theta\cos\theta_0\cos(\bar{\omega}_{\rm eff}T),\nonumber\\
\bar{\bf M}^{\rm C}_y &=&  \sin\Delta\theta\sin(\bar{\omega}_{\rm eff}T),\\
\bar{\bf M}^{\rm C}_z &=& \cos\Delta\theta\cos\theta_0 +  \sin\Delta\theta\sin\theta_0\cos(\bar{\omega}_{\rm eff}T),\nonumber
\eea

\begin{figure}
\includegraphics[width=3.3in]{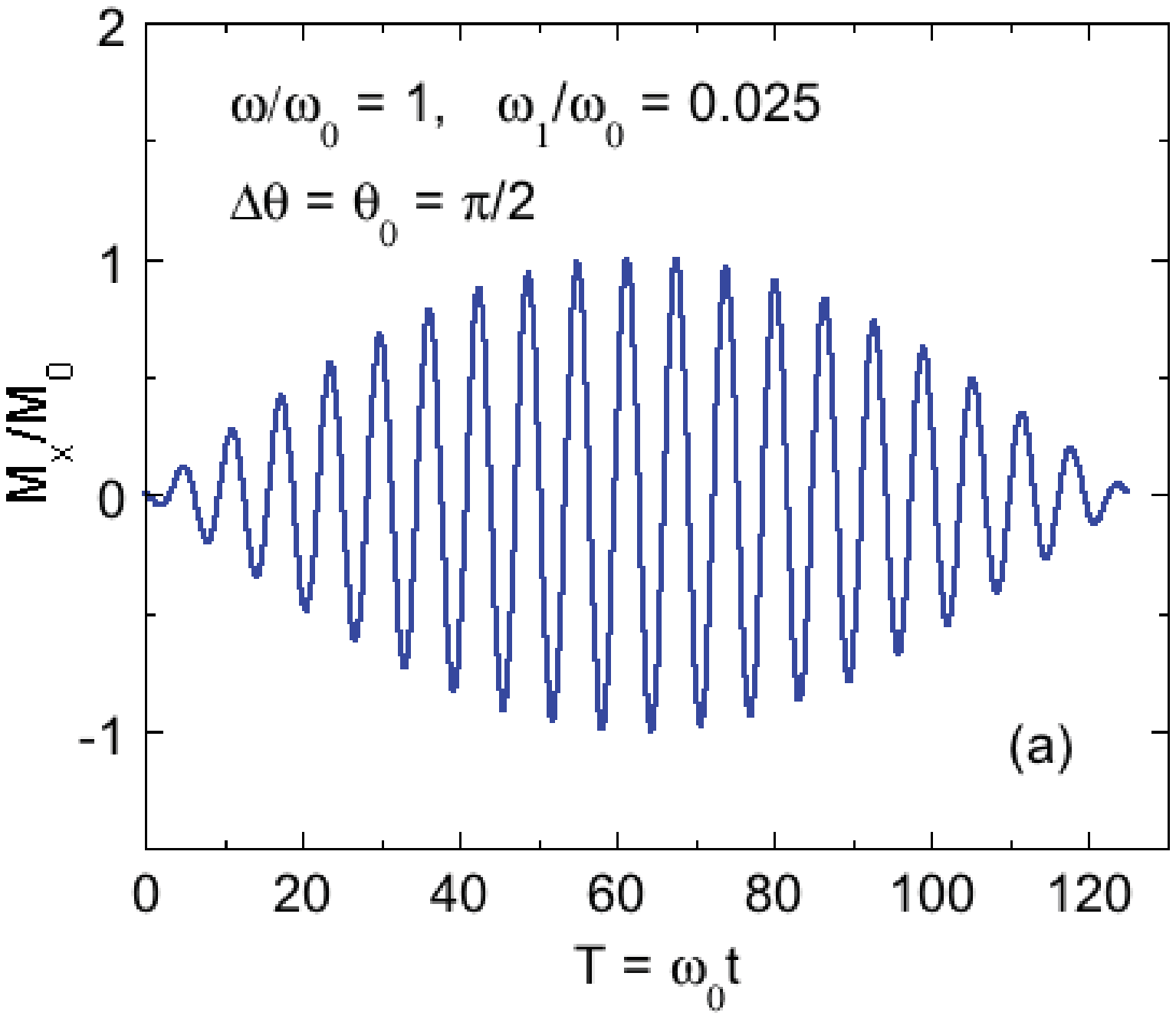}\vspace{0.3in}
\includegraphics[width=3in]{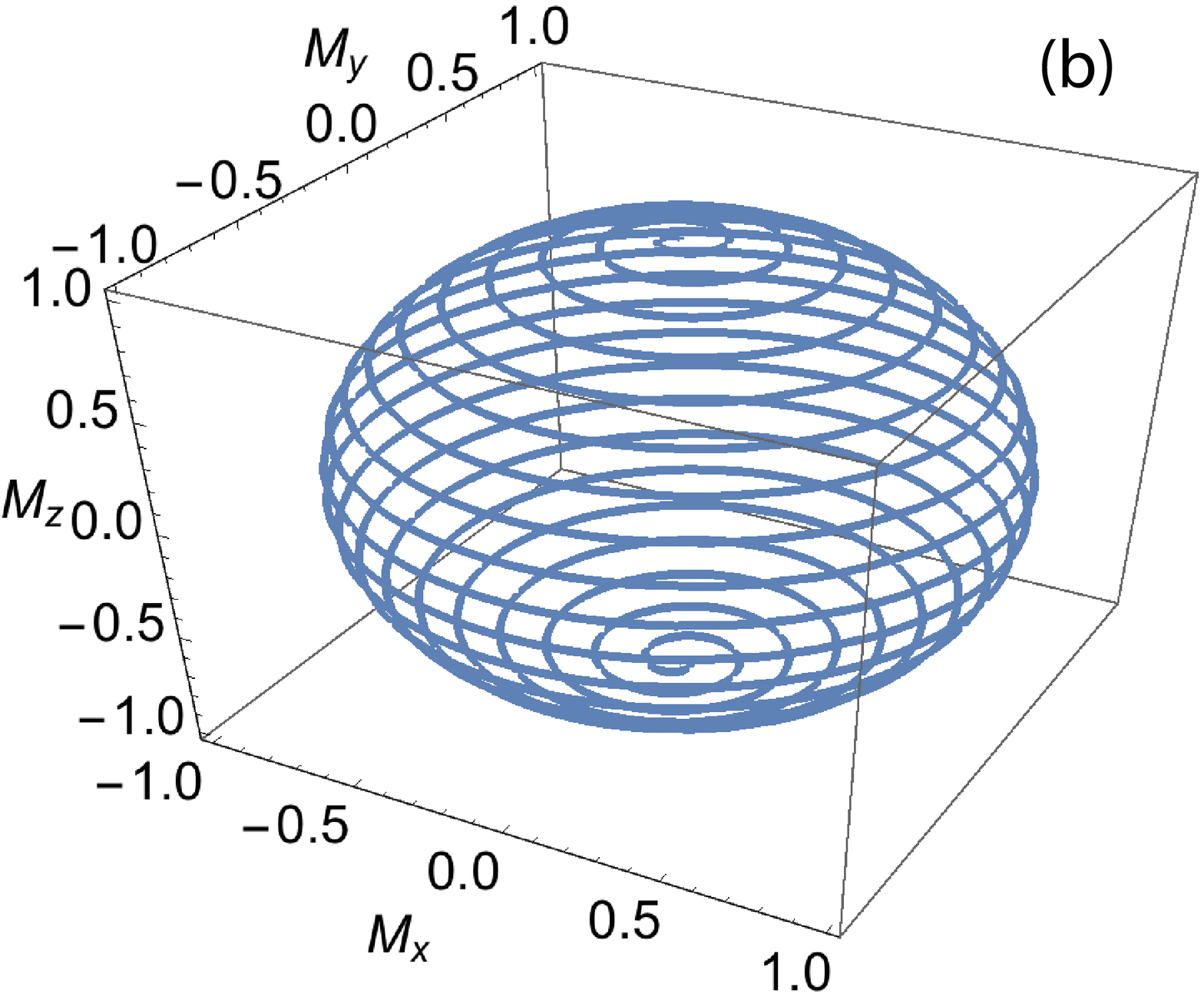}
\caption {These plots are for $\bar{\omega} = 1$, $\bar{\omega}_1 = 0.025$, and $\Delta\theta = \theta_0=\pi/2$~rad.  (a)~$\bar{M}_x \equiv M_x/M_0$ versus reduced time $T = \omega_0 t$.  (b)~Three-dimensional parametric plot of $\bar{M}_z$ versus $\bar{M}_x$ and $\bar{M}_y$  with~$T$ as the implicit parameter.  The plots in (a) and~(b) are for a time $T = 0$ to $\pi/\bar{\omega}_{\rm eff}$ (one-half period of $\bar{\omega}_{\rm eff})$.}
\label{Fig:muxVsT_w1.025_w1_DqQ0_0.5Pi}
\end{figure}

\begin{figure}
\includegraphics[width=2.9in]{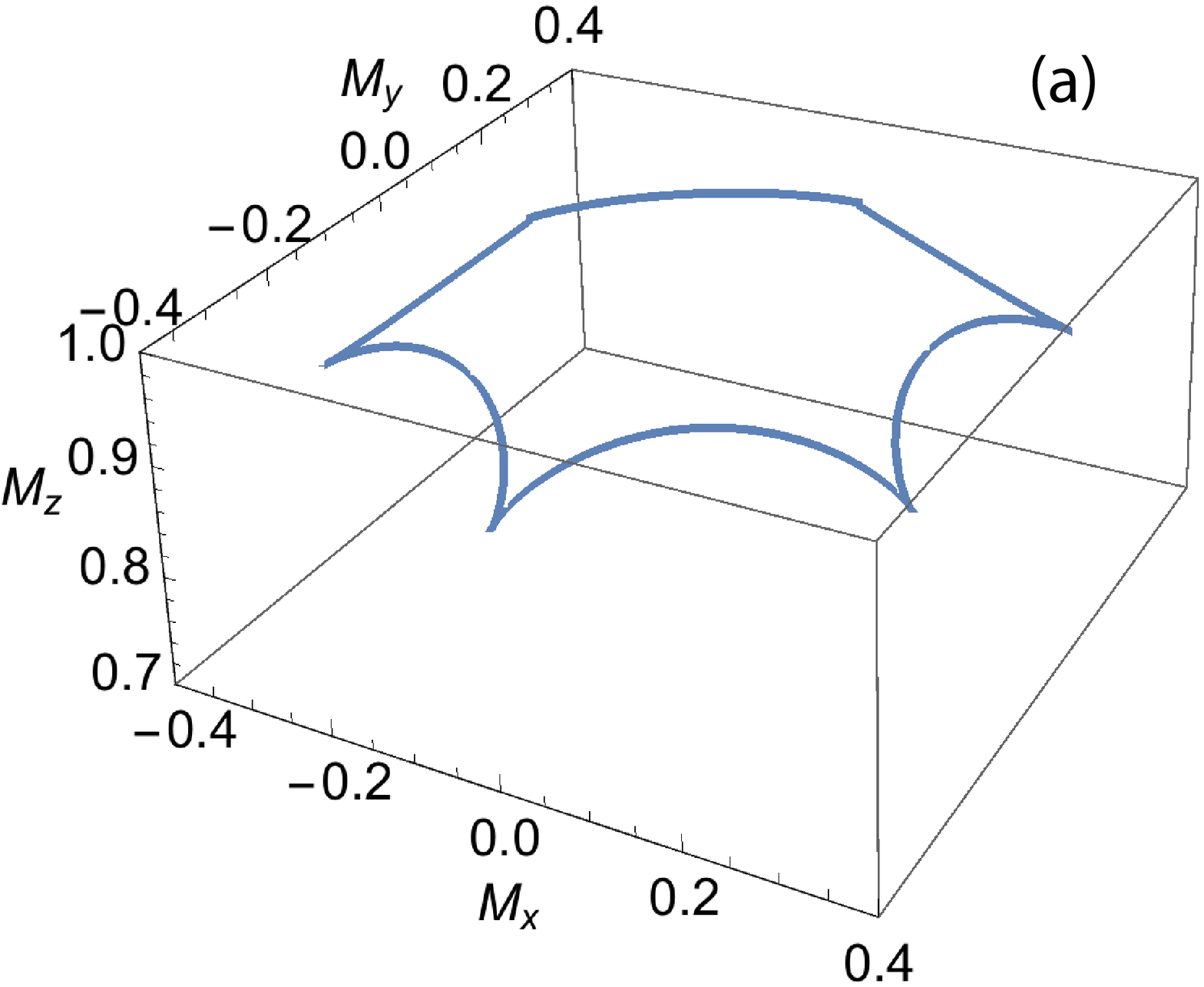}\vspace{0.3in}
\includegraphics[width=3.in]{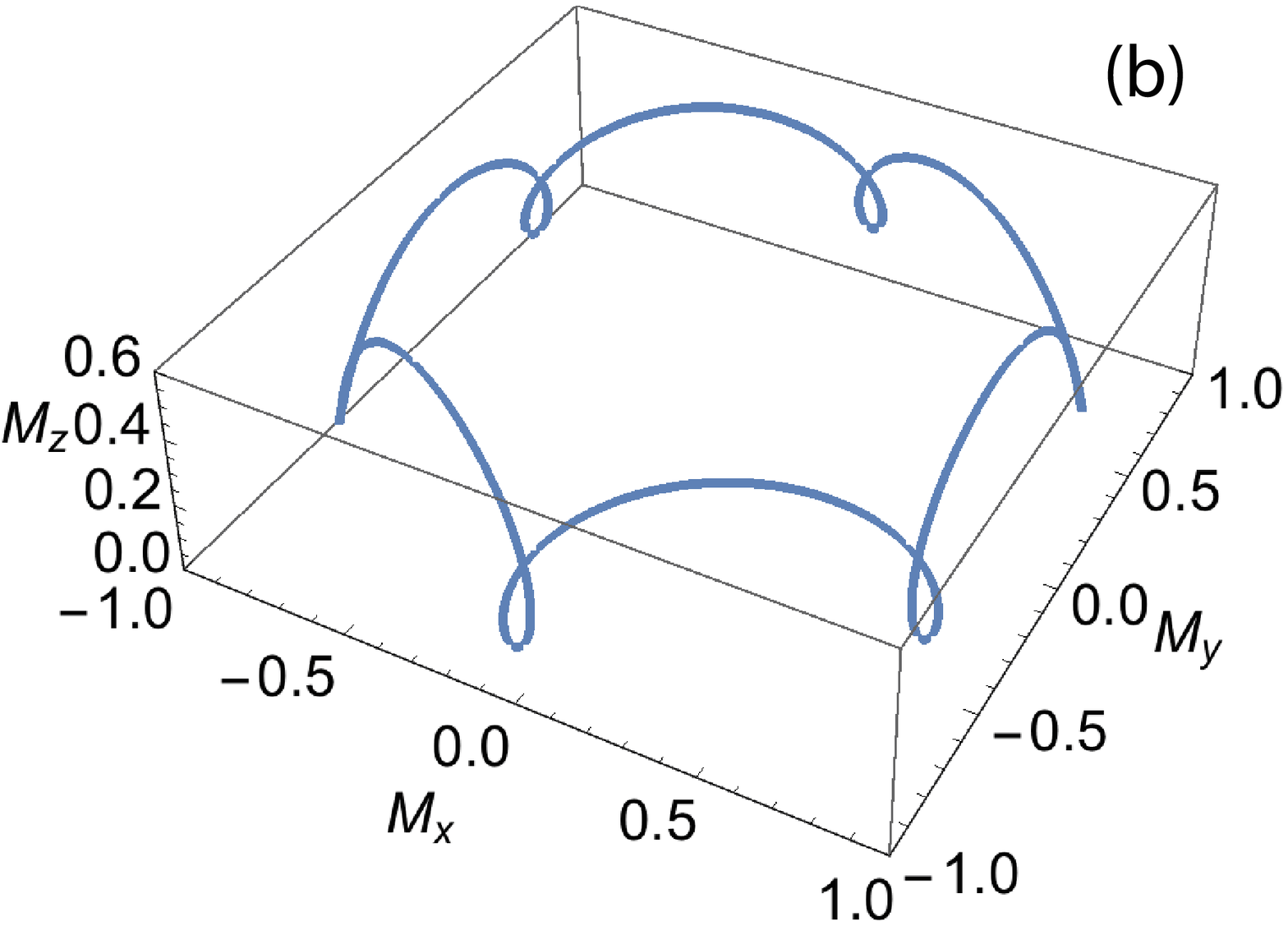}\vspace{0.3in}
\includegraphics[width=3.in]{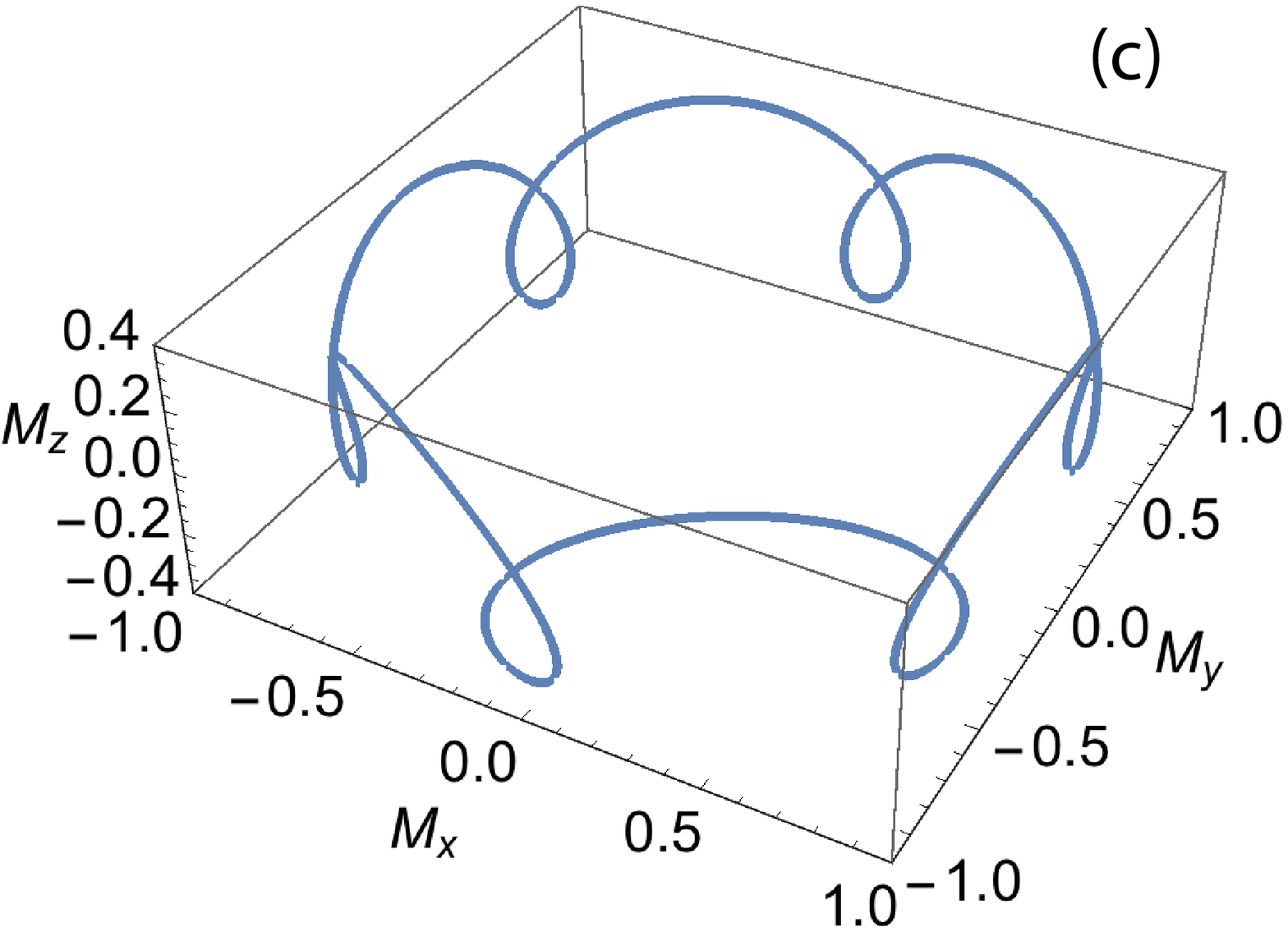}\vspace{0.3in}
\caption {3D parametric plots of $\bar{M}_z$ versus $\bar{M}_x$ and $\bar{M}_y$ with $T$ as the implicit parameter according to Eqs.~(\ref{Eqs:Mfinal}) and~(\ref{Eq:baromega1}) for $n=6$ and (a)~$\bar{\omega} = 0.15$, $\bar{\omega}_1 = 0.296$, and $\theta_0 = 2\Delta\theta = 0.335$~rad,  (b)~$\bar{\omega} = 0.3$, $\bar{\omega}_1 = 1.658$, and $\theta_0 = 2\Delta\theta = 1.171$~rad, and (c)~$\bar{\omega} = 1$, $\bar{\omega}_1 = 6$, and $\theta_0 = 2\Delta\theta = \pi/2$~rad.  Panel~(a) shows epi-cycloid behavior, whereas panels~(b) and~(c) increasingly show epi-prolate cycloid behavior.}
\label{Fig:3DMvsT_n6_omega0.15}
\end{figure}

Finally, rotating ${\bf M}^{\rm C}$ about the $z$~axis by a positive angle $\bar{\omega}T$ to obtain the precessing magnetization ${\bf M}(T)$ in Fig.~\ref{Fig:Undamped_Precession} gives
\bea
\bar{\bf M}_x &=&  \left[\cos\Delta\theta \sin\theta_0 - \sin\Delta\theta \cos\theta_0 \cos(\bar{\omega}_{\rm eff}T)\right]\cos(\bar{\omega}T)\nonumber\\
&& -\ \sin\Delta\theta \sin(\bar{\omega}T) \sin(\bar{\omega}_{\rm eff}T),\nonumber\\
\bar{\bf M}_y &=&  \left[\cos\Delta\theta \sin\theta_0 - \sin\Delta\theta \cos\theta_0 \cos(\bar{\omega}_{\rm eff}T)\right]\sin(\bar{\omega}T)\nonumber\\
&& +\ \sin\Delta\theta \cos(\bar{\omega}T) \sin(\bar{\omega}_{\rm eff}T),\nonumber\\
\bar{\bf M}_z &=& \cos\Delta\theta\cos\theta_0 +  \sin\Delta\theta\sin\theta_0\cos(\bar{\omega}_{\rm eff}T).\label{Eqs:Mfinal}
\eea

There are many combinations of $\bar{\omega}_1$, $\bar{\omega}$, and $\Delta\theta$ that can be considered.  Here we discuss a few representative cases.  In actual ESR experiments one usually has $\bar{\omega}_1\ll 1$ (unsaturated condition), but we need not be restricted to this inequality here.  Shown in Fig.~\ref{Fig:muxVsT_w1.025_w1_DqQ0_0.5Pi} are plots for $\bar{\omega}_1\ll 1$ at resonance (where $\bar{\omega}=1$), with $\bar{\omega}_1=0.025$ and $\Delta\theta = \theta_0=\pi/2$.  Figure~\ref{Fig:muxVsT_w1.025_w1_DqQ0_0.5Pi}(a) shows $\bar{M}_x$ versus reduced time~$T=\omega_0 t$ for one-half period of the effective frequency~$\bar{\omega}_{\rm eff}$, and a 3D plot of $\bar{M}_z$ versus $\bar{M}_x$ and~$\bar{M}_y$ with $T$ as the implicit parameter is shown for the same time period in Fig.~\ref{Fig:muxVsT_w1.025_w1_DqQ0_0.5Pi}(b).  The values of $\theta_0$ and~$\Delta\theta$ were chosen to give the initial conditions $M_x=M_y=0$ and $\bar{M}_z=1$ at $T=0$, so one can follow the path of $\bar{\bf M}$ versus time starting from $T=0$ at the top to $T = \pi$ at the bottom of Fig.~\ref{Fig:muxVsT_w1.025_w1_DqQ0_0.5Pi}(b).  During the second half of the period the path rotates with the same chirality upward on the spherical surface until the initial position is reached.

Of more interest to the present paper is the behavior of the path when $\bar{\omega}_1$ becomes appreciable compared to the resonant frequency $\bar{\omega}=1$, which is termed the condition for ``saturation'' in the field of ESR, a condition that is usually avoided in practice.  A commensurate value of $\bar{\omega}_{\rm eff}$ occurs when $\bar{\omega}_{\rm eff}/\bar{\omega}$ is an integer $n\geq 1$.  This is termed commensurate because a 3D plot of $\bar{M}_z$ versus $\bar{M}_x$ and $\bar{M}_y$ for such values of $\bar{\omega}_{\rm eff}$ and $\bar{\omega}$ overlaps after each period of reduced time $T = 2\pi/\bar{\omega}_{\rm eff}$.  Using Eq.~(\ref{Eq:omegaeff}), the equality $\bar{\omega}_{\rm eff}/\bar{\omega} = n$ for any value of $n> 1$ yields
\be
\bar{\omega}_1 = \sqrt{\bar{\omega}\left[(n^2 - 1)\bar{\omega} + 2 \right] - 1}.
\label{Eq:baromega1}
\ee

Shown in Fig.~\ref{Fig:3DMvsT_n6_omega0.15} are 3D parametric plots of $\bar{M}_z$ versus $\bar{M}_x$ and $\bar{M}_y$ with $T$ as the implicit parameter according to Eqs.~(\ref{Eqs:Mfinal}) and~(\ref{Eq:baromega1}) for $n=6$ and other parameters listed in the figure caption.  Since $n=6$, one sees a sixfold periodic rotational behavior in each of the three panels.  Figure~\ref{Fig:3DMvsT_n6_omega0.15}(a) shows a case where the path of $\bar{\bf M}$ is an epi-cycloid, whereas Figs.~\ref{Fig:3DMvsT_n6_omega0.15}(b) and~\ref{Fig:3DMvsT_n6_omega0.15}(c) show increasingly epi-prolate-cycloid behaviors.  With increasing values of $\bar{\omega}$, $\bar{\omega}_1$ increases from 0.296 in panel~(a) to~6 in panel~(c), and $\theta_0$ also increases from 0.335~rad in panel~(a) to $\pi/2$~rad in panel~(c).  These paths and those in Fig.~\ref{Fig:muxVsT_w1.025_w1_DqQ0_0.5Pi}(b) are three-dimensional angular variations with time at constant radius~$\bar{M}=1$ which are to be contrasted with the cycloidal paths of a moon about a star in Fig.~\ref{Fig:ycloid_alph0.1_ronR0.05} where the two-dimensional cycloidal variations with time are in the radial distance of the moon from the star.

\section{\label{Sec:Summary} Summary}

In this paper, the cycloidal paths of a point P in several physical situations of practical interest are studied.  The parametric equations for the path of~P a distance~$r$ from the axis of a disk with radius~$R$ that is rolling without slipping is derived from the superposition of the translational motion of the center of mass of the disk and the rotational motion of P about the center of mass.  As previously known, a cycloid path and also curtate and prolate cycloid paths are found for $r=R$, $r < R$ and $r>R$, respectively.  In these cases the cycloid is described parametrically in terms of the Cartesian $x$ and $y$ coordinates of P as a function of time as the implicit parameter.  The same forms of cycloidal paths are obtained during rolling with frictionless slipping, but where the time dependence of the sinusoidal Cartesian coordinates of the point P is modified.  The parametric equations versus time are obtained for the orbit with respect to a star of a moon in a circular orbit about a planet that is in a circular orbit about a star, where the orbits are coplanar, using the same approach as for the rolling disk.  Here the radial distance of the moon from the star is the parameter showing cycloidal paths.  Finally, we show that cycloidal paths of the magnetization vector versus time can occur during undamped electron-spin resonance if the amplitude $H_1$ of the microwave magnetic  field is an appreciable fraction of the magnitude $H$ of the applied static magnetic field.

\acknowledgments

This research was supported by the U.S. Department of Energy, Office of Basic Energy Sciences, Division of Materials Sciences and Engineering.  Ames Laboratory is operated for the U.S. Department of Energy by Iowa State University under Contract No.~DE-AC02-07CH11358.

%\clearpage

%\acknowledgments

%\clearpage

\end{document}